\documentclass[aps,twocolumn,prb,showpacs,floatfix,superscriptaddress]{revtex4}
\usepackage{graphics,amssymb,amsmath,epsfig}
\begin{document}
\setcounter{page}{1}
\title[]{ Coarsening Dynamics in a Two-dimensional XY model with Hamiltonian Dynamics}

\author{Kyo-Joon Koo}
\affiliation{Department of Physics, University of Suwon, Kyonggi-do 445-743, Korea}

\author{Woon-Bo Baek}
\affiliation{Department of Mechatronics, Dong-Eui University, Busan 614-714, Korea}

\author{Bongsoo Kim}
\affiliation{Department of Physics, Changwon National University, Changwon 641-773, Korea}

\author{Sung Jong Lee}
\affiliation{Department of Physics, University of Suwon, Kyonggi-do 445-743, Korea,}
\affiliation{School of Computational Sciences, Korea Institute for Advanced Study, Seoul 130-722, Korea}

%\author{Kyo-Joon \surname{Koo}}
%\affiliation{Department of Physics, University of Suwon, Kyonggi-do 445-743, Korea}
%\author{Woon-Bo \surname{Baek}}
%\affiliation{Department of Mechatronics, Dong-Eui University, Busan 614-714, Korea}
%
%\author{Bongsoo \surname{Kim}}
%\affiliation{Department of Physics, Changwon National University, Changwon 641-773, Korea}
%
%\author{Sung Jong \surname{Lee}}
%\affiliation{Department of Physics, University of Suwon, Kyonggi-do 445-743, Korea,}
%\affiliation{School of Computational Sciences, Korea Institute for Advanced Study, Seoul 130-722, Korea}

%\date[]{Revised Sept. 4 2006}

\begin{abstract}
 We investigate the coarsening dynamics in the two-dimensional
Hamiltonian XY model on a square lattice, beginning with a random state
with a specified potential energy and zero kinetic energy. 
Coarsening of the system proceeds via an increase in the kinetic energy and 
a decrease in the potential energy, with the total energy being conserved. 
We find that the coarsening dynamics exhibits a consistently superdiffusive
growth of a characteristic length scale as 
$L(t) \sim t^{1/z}$ with $1/z  > 1/2$ (ranging from $0.54$ to $0.57$ for typical
values of the energy in the coarsening region).
Also, the number of point defects (vortices and antivortices) decreases
as $N_V(t) \sim  t^{-\phi_V}$, with $\phi_V$ ranging between $1.0$
and $1.1$. On the other hand, the excess potential energy decays
as $ \Delta U \sim t^{-\phi_U}$, with a typical exponent of $\phi_U \simeq 0.88$,
which shows deviations from the energy-scaling relation. 
The spin autocorrelation function exhibits a peculiar time 
dependence with non-power law behavior that can be fitted well by an exponential
 of logarithmic power in time.
We argue that the conservation of the total Josephson (angular) momentum plays
a crucial role for these novel features of coarsening in the Hamiltonian {\em XY} model.
\end{abstract}


\maketitle

%\begin{references}
%%\begin{thebibliography}{}
%\bibitem{R1} G. D. Hong and Jang Lee, J. Korean Phys. Soc. {\bf 44}, L1 (2004).
%\bibitem{R2} S. H. Yoon, J. Korean Phys. Soc. {\bf 41}, 37 (2002).
%%\end{thebibliography}
%\end{references}

%\documentclass[aps,twocolumn,prb,showpacs,floatfix,superscriptaddress]{revtex4}
%\usepackage{graphics,amssymb,amsmath,epsfig}
%%\documentstyle[aps,prl,epsfig,multicol]{revtex}
%\begin{document}

%relaxation \sep superconducting arrays \sep frustration
%\pacs{74.50+r, 67.40.Fd}

\section{INTRODUCTION}

%Fundamental assumption behind equilibrium statistical physics is
%the so called ergodic hypothesis stating that, for a dynamic system
%with large degrees of freedom, the molecular chaos ensures that the
%time average of a physical measurable can be replaced by ensemble average
%with some thermodynamic parameter \cite{review}..
In recent years, there have been research efforts toward
investigating the statistical mechanical behavior of lattice spin
systems by solving directly the Hamiltonian equation of motion
associated with the lattice spin system of interest
\cite{casetti_hamil_review}. Especially, equilibrium phase
transitions have been found to be associated with the existence of
topological change in the dynamics, as manifested in the
distributions of Lyapunov exponents \cite{caian1,
caian2,leoncini,cerruti,latora,zheng_ising,zheng_ising_ordering,
kockel_ising_ordering_comment}.

%Hamiltonian of the system is defined as
%
%\begin{eqnarray}
%H & =  & \sum_{i} \frac{p_i^2}{2} - J \sum_{<ij>} (1- \cos(\theta_i - \theta_j )  \\
%  & = & \sum_i  { {\dot{\theta}_i^2} \over {2}} - J \sum_{<ij>} (1- \cos(\theta_i - \theta_j)
%\end{eqnarray}
%

In relation to these interests, we here investigate the coarsening
dynamics of the {\em XY} model on a square lattice based on Hamiltonian
dynamics. In conventional studies of phase ordering and coarsening
dynamics in a statistical system, the system is quenched from the
random disordered state to a low-temperature state below the
transition temperature\cite{ordering_review, bray_review}. In
computational studies, one employs Monte Carlo or Langevin dynamics
methods for simulating the coarsening processes. In typical
situations, the average length scale $L$ of ordered domains grows in
time as a power law $L(t) \sim t^{1/z}$, where the growth exponent
$1/z$ depends on the dimension of the space and the dimension of the 
relevant order parameter, in addition to the conserved or nonconserved
nature of the latter in the relaxation dynamics\cite{bray_review}. Usually,
characteristic topological defects, such as point vortices or domain
walls, are generated in the initial disordered state, depending on the
dimension of the order parameter, and the annihilation of these
defects gives the main mechanism of coarsening and phase ordering in
the system. The observed self-similarity of these coarsening systems
at different time instants is usually represented by the so-called
{\em dynamic scaling hypothesis} of the equal-time spatial
correlation function of the order parameter \cite{bray_review,bray_review_2}.

 In equilibrium, the {\em XY} model exhibits a Kosterlitz-Thouless (KT)
 transition at $T_{KT}$ due to the unbinding of vortex-antivortex pairs
\cite{BKT}. Below $T_{KT}$, the system has a quasi-ordered phase
that is characterized by a power-law decay of the order parameter
correlation function for long distances.  The critical exponent
governing the power-law decay decreases {\it continuously} down to
zero temperature: the system is critical at equilibrium for all
temperatures below $T_{KT}$.

After some efforts of various groups
\cite{loft,toyoki3,mondello,bray,yurke,blundell,jrl,rojas,bray2,bray3,ying},
as well as some controversies, it is now agreed that, in the phase
ordering dynamics of the {\em XY} model via Monte Carlo simulations or
Langevin dynamics methods, the growing length scale exhibits a
logarithmic correction\cite{wilczek} to diffusive growth as
$L_{MC}(t)\sim [t/(\log t)]^{1/5}$. Here, the logarithmic correction
can be attributed to a logarithmic divergence (in the system size)
of the effective friction constant of a moving vortex in the dissipative
dynamics with a non-conserving order parameter. In a related
simulational work on the coarsening dynamics of superconducting
juction arrays based on the dynamics of the resistively-shunted junction
(RSJ) model, it was revealed that there is no logarithmic correction.
This could be understood in terms of a finite friction constant of 
a moving vortex in the limit of large system size, which is due to 
the particular type of dissipative coupling in Josephson junction arrays
\cite{jj_ordering}.

% (spin-rotor) dynamics.
One of the questions we would like to address in this work is
whether we would find similar features as mentioned above in the
coarsening of the {\em XY} model based on Hamiltonian dynamics. 
Unlike the case of conventional coarsening based on dissipative
dynamics, almost no works exist on coarsening via Hamiltonian
dynamics, except for the case of Ising-type systems. 
In the simpler case of the Ising-type $\phi^4$ model on a two-dimensional 
square lattice with
nonconserved order parameter, coarsening via Hamiltonian dynamics
was investigated by Zheng \cite{zheng_ising_ordering}. He contends
that the growth exponent is different from the usual diffusive
growth of $1/z = 1/2$ obtained in the traditional phase ordering.
This result, however, was later disputed by others as a crossover
effect \cite{kockel_ising_ordering_comment}, which, we think, is
still controversial because the asymptotic dynamic features are not
 completely settled yet. Now, in the case of the
{\em XY} model with Hamiltonian dynamics, it is of some interest to
investigate whether any new features exist in the
non-equilibrium coarsening processes. Due to the Hamiltonian
dynamics nature of our system, the total energy is conserved, and we
specify the initial states by their total energy divided by the
system size, i.e., the per-site energy or the energy density. 
These initial states with specified energies are prepared with 
special Monte Carlo algorithms to be explained in section II. 
In order to make an analogy to ordinary coarsening in dissipative
 dynamic relaxation more apparent, the initial rotational velocities of 
the rotors are taken to be zeroes.

Since we begin with zero kinetic energy (zero rotational velocities 
for all spins), the Hamiltonian dynamics of the system will generate 
kinetic energy taken from the potential energy. As time proceeds,
we can expect that (for a per-site energy that is low enough to correspond
to a low-temperature quasi-ordered phase), the system will evolve toward 
some equilibrium stationary dynamic state, which can be considered as
 the thermal equilibrium state in the case of purely dissipative dynamics.
In other words, the kinetic energy that is generated during the 
course of coarsening process acts as a thermal bath for the system.
We investigate the time-dependent spin configuration for the system 
in terms of spatial ordering and relaxation of the vortex numbers, etc., 
in analogy to the conventional dissipative coarsening systems based
 on Monte Carlo or Langevin dynamics.

 We find that the equal-time spatial correlation functions approximately satisfy
critical dynamic scaling
\begin{equation} \label{eq:dynamic_scaling}
C(r,t)= r^{-\eta(T)} g(r/ L(t)),
\end{equation}
with a spatial correlation exponent $\eta (T)$ (which is
increasing in $T$) and a growing length scale $L(t)$ that grows typically
as $L(t) \sim t^{1/z} $, where $z$ is the dynamic exponent. 
In the late time region, we find that the
length scale $L(t)$ grows with an exponent $1/z$ that is 
larger than the diffusive exponent $1/2$. Closely related to this
growing length scale is the average separation between vortices $L_{V}
(t)$, which can be derived from the decay of the number of vortices
$N_V (t) $, which behaves as $N_V (t) - N_V (\infty) \sim t^{-\phi_V }$.
For unform distribution of vortices, we expect $ L(t)
\sim L_V (t) \sim (N_V (t))^{-1/2} \sim t^{-\phi_V /2}$ in the limit where
$N_V (\infty)$ is negligible (which appears to be valid approximately
in the low-energy region of $E < 0.5$). We see that the vortex density 
relaxations exhibit (approximately) the expected behavior with the 
exponents $\phi_V $ being larger than unity.

We find also that the relaxation of the potential energy exhibits a power-law
decay $ \Delta U \equiv U(t)-U_{\infty} \sim t^{-\phi_U}$ toward 
a long-time equilibrium value, with the exponents taking values approximately 
in the range $ 0.86 < \phi_U  < 0.90$ (in typical cases, $\phi_U \simeq 0.87 \sim 0.88$).
This shows a considerable deviation from the energy-scaling relation relating 
the excess energy and the growing length scale \cite{bray-ruten}.     
 Another interesting feature of the coarsening dynamics in the Hamiltonian
 {\em XY} model is that the spin autocorrelation exhibits a peculiar 
non-power-law behavior, which can be reasonably well fitted
by $ A(t) \sim  A_0 \exp [ -b (\ln (t) )^{\gamma} ]$,
 with $b$ ranging from $0.24$ to $0.4$ and the exponent $\gamma $ from
$1.5$ to $1.7$.

In order to understand the microscopic mechanism of these coarsening features,
including the notable superdiffusive growth exponent, that are quite
distinct from the cases of purely dissipative dynamics,  
we note that one conserved quantity exists in this Hamiltonian
model system.  That is, the total angular momentum (or Josephson
angular momentum) is conserved in the Hamiltonian dynamics. 
Due to the conservation of angular momentum, propagating spin
wave modes appear \cite{das-rao1, das-rao2}. We strongly believe that the 
existence of propagating spin wave modes due to the conservation of 
angular momentum is the main cause of the superdiffusive nature of the 
coarsening and of the other unique features in this system. In order to 
obtain an effective friction constant of a vortex, we could apply a collective
variable approach. However, this method gave us a vanishing effective
friction constant to the lowest order in the velocity of a
vortex. Higher-order friction can be argued to increase the growth 
exponent $1/z$ from the diffusive value of $1/2$.
Further work is necessary to reach a full understanding of the
main features of the coarsening in the Hamiltonian {\em XY} model in two
dimensions.

\bigskip

\section{The Hamiltonain XY model and Simulation Methods}

\bigskip

% The Ginzburg-Landau Hamiltonian of $O(2)$ model is given by
%
%\begin{equation}
%H=\int d^2 r [{1 \over 2}(\nabla \vec{\phi})^2 +
% {1 \over 4}(\vec{\phi}^2-1)^2]
%\end{equation}
%where $\vec{\phi}$ is a two component real vector field:
%$\vec{\phi}=(\phi_1, \phi_2)$.
%The time evolution of the model is assumed to be governed by
%the model A dynamics appropriate to the non-conserved order
%parameter, of the form
%
%\begin{equation}
%{\partial \phi_{\alpha} \over \partial t} = -{ \delta H
% \over \delta \phi_{\alpha} } + \zeta_{\alpha} (\vec{r},t)
% =\nabla^2 \phi_{\alpha} +(1-\vec{\phi}^2)\phi_{\alpha}+
%  \zeta_{\alpha} (\vec{r},t),
% \;\;\; \alpha =1,2
%\end{equation}
%where the thermal noise $\zeta_{\alpha}(\vec{r},t)$ is white gaussian with
%zero mean and with variance satisfying the detailed balance
%at temperature $T$.
%
%\begin{equation}
%<\zeta_{\alpha}  (\vec{r},t) \zeta_{\beta} (\vec{r}^{\prime},t^{\prime})>=
%2 k_B T \delta_{\alpha \beta}\delta (\vec{r}-\vec{r}^{\prime})
%\delta (t-t^{\prime}).
%\end{equation}

The potential energy function of the hard-spin {\em XY} model on
a square lattice is given by
\begin{equation} \label{eq:potential}
U= J\sum_{<ij>} (1- \cos(\theta_{i}-\theta_{j}) ),
\end{equation}
where $J$ is the interaction strength, $\theta_i$ is the phase angle
of the spin at site $i$ and the sum is over nearest neighbor pairs. 
% $\vec{S}_i \equiv (\cos \theta_i , \sin \theta_i )$
The corresponding Hamiltonian dynamic equation can be obtained
by adding the kinetic term: 
\begin{equation} \label{eq:Hamiltonian1}
H  =   \sum_i   \frac{m_i^2}{2} + J \sum_{<ij>} (1- \cos(\theta_i - \theta_j ), 
%  & = & \sum_i  {\frac{\dot{\theta}_i^2}{2}} + J \sum_{<ij>} (1- \cos(\theta_i - \theta_j)
\end{equation}
where $m_i = \dot{\theta}_i$ represents the canonical momentum associated with
the angular variable $\theta_i$ of planar spins at site $i$. 
This model appears naturally in easy-plane ferromagnets or superfluid
helium with the conserved variable $m$ corresponding to the $z$-component of 
the spins or the density of the superfluid respectively in the limit of
negligible themal noise (e.g., negligible phonon effect) 
\cite{hohen_halperin,nelson-fisher}.             
  
Previous works \cite{leoncini} have shown that a Kosterlitz-Thouless (KT) transition
occurs at the value of potential energy around $U_{KT}\simeq 1.0 $. 
Our computational scheme in this work performs Newtonian dynamic simulations 
based on the above Hamiltonian with initial states that take zero kinetic energy 
and specified potential energy values but are otherwise random. Specified potential 
energy values are chosen from the region that corresponds to below or near
the KT transition energy ($U_{KT} \simeq 1.0$).

Unlike the case of dissipative systems with Langevin dynamics or 
Monte Carlo dynamics where dissipative relaxation is incorporated 
in a natural way, here in our case of Hamiltonian dynamics, the 
total energy (sum of kinetic and potential energy) is
conserved. However, since we begin the Hamiltonian dynamics with zero
kinetic energy with a given specified potential energy, we expect that 
the system will evolve in such a way as to increase (on the average)
the kinetic energy and decrease the potential energy. Eventually, in the long-time
limit, the system will reach a certain kind of equilibrium dynamic state 
where the average potential energy and the kinetic energy do not vary
anymore. Of course, in terms of phase-ordering dynamics in dissipative 
systems, a system quenched to a symmetry-broken phase will never reach 
complete equilibrium in a finite time period.

Therefore, we can consider the kinetic part of the system acting as
a kind of thermal heat bath that absorbs the "dissipated heat"
coming from the decreasing potential energy. We can, thus, investigate
the time-dependent spin configuration for the system in terms of
spatial ordering and relaxation of vortex numbers, etc., in analogy
to the Monte Carlo or Langevin relaxation dynamics. We generate the
initial states with specified potential energies in the following
manner: To begin with, we take a random spin configuration. Then we
employ a Monte Carlo annealing algorithm such that the spin
configurations with potential energies that are closer to the target
energy are preferentially accepted. In other words, we steer the
initial state to some target spin configuration such that we obtain
a state whose potential energy is equal to a specific energy; 
otherwise, the spin configuration is random.

% We might call this method a guided annealing method.

%Langevin dynamics
%
%\begin{equation}
%{\partial \theta_{i} \over \partial t} = -{ \delta H
% \over \delta \theta_{i} } + \zeta_{i} (t), \;\;\; i=1,\cdots, N^2
%\end{equation}
%where
%the thermal noise $\zeta_i (t)$ satifies
%\begin{equation}
%<\zeta_{i} (t) \zeta_{j} (t^{\prime})>=2 k_B T \delta_{ij}
%\delta (t-t^{\prime}).
%\end{equation}
%Langevin dynamics Done

The Hamiltonian equation reads
%{\ddot{\theta}_{i} } & = & - \frac{\delta V}{\delta \theta_{i}}    \\
%    &      =  &  - J \sum_{\hat{n}} \sin (\theta_i -\theta_{i+\hat{n}}),  i=1, \cdots, N^2
\begin{eqnarray} \label{eq:hxy-eq}
{\dot{m}_{i} } & = & - \frac{\delta U}{\delta \theta_{i}} = -J \sum_{j} \sin (\theta_i -\theta_{j}), \\
{\dot{\theta}_{i}}   &    =  & m_{i}
\end{eqnarray}
where $j$ denotes the nearest neighbors of site $i$ and  $i$ takes values  $i=1, \cdots, N^2$.
For simplicity, we set the rotational inertia to be equal to unity and
put $J=1$. Here, we note that, in addition to the conservation of 
total energy, another conserved quantity, namely the total
(Josephson) angular momentum $\sum_{i} m_{i} \equiv \sum_{i}\dot{\theta}_{i}$ exists; This is
due to the invariance of the equations under global translations in the variable 
$\theta_{i}$, that is, under $\theta_{i} \rightarrow \theta_{i} + \alpha$.  

Equations~4 and 5 are numerically integrated in time using a second
order velocity-Verlet algorithm \cite{yoshida} with the time integration step of
$\Delta t = 0.01$, which conserves the total energy to within a ratio
of $10^{-4}$ up to the maximal time integration steps of around $10^6$. 
Periodic boundary conditions on both lattice directions are employed. The simulations are
carried out on square
lattices of dimensions up to $N\times N =1800 \times 1800$ with sample averages
of over $20$ to $40$ different initial configurations. A parallel computation 
is employed on Linux clusters with from $16$ to $32$ CPU's via a domain decomposition
method with high parallel efficiency.

The final results are obtained from averages over $10$ to $40$ different
random initial configurations.
The main quantities of interest are as follows:

(i)  Equal time spatial correlation function of the order parameter,

\begin{equation}
C (r,t)  =  {1 \over N^2} \left < \sum_{i} \cos (\theta_{i}(t) - \theta_{i+r}(t) \right >,
%& \equiv & {1 \over N^2} \left < \sum_{i} \vec{S}_{i}(t) \cdot \vec{S}_{i+r}(t) \right > \\
\end{equation}
where $< \cdots >$ denotes an average over random initial configurations.

(ii) Number of vortices $N_V (t)$ at time $t$, including both positive and
     negative vortices.

(iii) Time-dependent relaxation of the potential energy, $U(t)$,

\begin{equation}
 U(t) \equiv {1 \over N^2} \left < \sum_{<i,j>}
  (1- \cos (\theta_{i}-\theta_j ) ) \right >,
\end{equation}

(iv) Nonequilibrium spin autocorrelation function

\begin{equation}
{A} (t) \equiv {1 \over N^2} \left < \sum_{i}
\cos ( \theta_{i}(0) - \theta_{i}(t) ) \right >.
\end{equation}
The results of simulations will be presented in section III.

\bigskip

\noindent

\section{Simulation Results and Discussions}

\bigskip

Figure~1 shows snapshots of spin configurations for the case of $E=0.4$ at four
different time stages of coarsening. We can see that coarsening occurs via decay
of vortices and anti-vortices.
%%\bigskip
%%  [[[ FIGURE 1 ]]]
%%\bigskip
Figure~2(a) shows the equal-time spatial correlation for $E=0.4$.
The simplest way to analyze the data is to attempt the so-called critical dynamic scaling
\begin{equation} \label{eq:dynamic_scaling2}
C(r,t)= r^{-\eta(E)} g(r/ L(t)),
\end{equation}
with the spatial correlation exponent $\eta (E)$
and a length scale $L(t)$ growing typically as $L(t) \sim t^{1/z} $.
In our case, for a given time instant $t$, we determined $L(t)$ in such a way that
$ r^{-\eta(E)} C(r,t)|_{(r=L(t))} = g(r/L(t))|_{(r=L(t))} = g(1) = 0.4 $.
For a suitable choice of $\eta(E)$, we can obtain a good collapse of the rescaled correlation
functions $g(x)$, at least in the late-time region.
Shown in Fig.~2(b) is such an attempt at critical dynamic scaling.
We find that the equal-time spatial correlation functions exhibit a reasonable
critical dynamic scaling, at least, in the late-time region. However, in the early-time 
regime, we can see that it is not possible to collapse the correlations with any
reasonable values of $\eta$.

%%%%%%%%%%%%%%%%%%%%%%%%%%%%%%%%%%%%%%%%%%%%%%%%%%%%%%%%%%%%%%%%%%%%%%
%\begin{table}[h]\centering
%\begin{tabular}{cccccc} \hline \hline
%  $(m^2,g)$ &  & $t_1=40$ &   $t_1=80$ &  $t_1=160$ &  $t_1=320$     \\
%  (8.0,2.4) &        & 2.84(5)  &   2.83(6)  &  2.76(10)  &  2.79(12)\\
%  (6.0,5.4) &        & 2.75(2)  &   2.75(1)  &  2.70(1)   &  2.72(1)\\
%  (6.0,1.8) &        & 2.78(3)  &   2.76(1)  &  2.67(3)   &  2.57(4)\\
%            & L=1024 & 2.72(7)  &   2.69(9)  &  2.67(7)   &  2.66(4) \\
%            & $\epsilon=\epsilon_{0}+4/3$
%                     & 2.78(5)  &   2.74(8)  &  2.70(10)  &  2.69(10)\\ \hline \hline
%\end{tabular}
%\caption{The dynamic exponent $z$ estimated from scaling collapse
%of $C(r,t)$ in a time interval $[t_1,640]$. If not specified,
%the lattice size $L=512$ and the energy density is at its
%fixed points.
%}
%\label{t1}
%\end{table}

\begin{table}[h]\centering
\begin{tabular}{|c|cccc|} \hline \hline
      E &  $\eta$ ($N=1000$)   & $1/z$ ($N=1000$) & $\eta$($N=1800$) &  $1/z$ ($N=1800$)
                      \\
\cline{1-5}
   0.1 &  0.046(10)   &      0.520(12)   &   &             \\
\cline{1-5}
 0.2 & 0.065(16) & 0.540(15) &     0.09(2)  & 0.549(12)         \\
 \cline{1-5}
   0.3     & 0.104(16) &    0.566(10)    &     &              \\
%\hline \hline
   0.4  &  0.112(17)   & 0.565(12)    & 0.115(18) & 0.555(10)               \\
\cline{1-5}
  0.5   & 0.120(18)     & 0.549(13) &  &    \\
\cline{1-5}
 0.6 & 0.135 (19) & 0.549(13) &  &   \\
\cline{1-5}
0.7   &   &     &0.156(21) &0.560(14)    \\
\cline{1-5}
0.8   &   &     &0.170(22) &0.565(14)    \\
\cline{1-5}
1.0   &   &     &0.235(20) &0.62(25)    \\
\hline\hline
\end{tabular}
\caption{Correlation exponent $\eta$ and the growth exponent $1/z$.}
%\protect\cite {fis88,bra94}.
\label{t1}
\end{table}

%%%%%%%%%%%%%%%%%%%%%%%%%%%%%%%%%%%%%
%%\bigskip
%%   [[[ FIGURE 2 ]]]
%%\bigskip

In the late-time region, we find that the length scale $L(t)$ grows
with an exponent $1/z$ that is larger than the diffusive
exponent $0.5$. For example, in the case of $E=0.4$, we obtain $1/z
\simeq 0.555 \pm 0.010$, which is definitely larger than $1/2$
(Fig.~2(c), Fig.~2(d), and Table~1). Note that the power law region shows up at $t
> t_{P} \approx 50$, which is a considerably late-time stage.

Closely related to this growing length scale is the average
separation between vortices, $L_V (t)$, which can be derived from the
decay of the number of vortices, $N_V (t)$. Shown in Fig.~3(a) are the
number of vortices versus time, which behaves as $N_V (t) - N_V
(\infty) \sim t^{-\phi_V }$, where $N_V (\infty)$ denotes the
limiting value of the vortex number at equilibrium.
%is negligible for energy $E \leq 0.5$. 
At low temperature, the equilibrium vortex density becomes vanishingly
small. We regard $N_V (\infty)$ as a fitting parameter such that the
relaxation of the excess vortex number obeys the best power-law
behavior. We find that the relaxation of the vortex number exhibits
an excellent power-law behavior except for the early-time region
where smooth plateau is seen before crossing over to the late-time 
power-law region. The exponent $\phi_V $ takes values in the range from
$1.0$ to $1.10$, depending on the system size and the energy, except for the
limit of higher per-site energy. 
 For example, in the case of $E=0.4$ with
$L=1800$, we get $\phi_V \simeq 1.055$. Table~2 and Figure~3(c) shows 
the $E$-dependence of the $\phi_V$.

\begin{table}[h]\centering
\begin{tabular}{|c|cccc|} \hline \hline
      E &  $\phi_V$ ($N=1000$) & $\phi_U$ ($N=1000$) & $\phi_V$($N=1800$) &  $\phi_U$ ($N=1800$)
                      \\
\cline{1-5}
   0.1 &  1.001(6)   &     0.870(15)   &   &             \\
\cline{1-5}
 0.2 & 1.059(8) & 0.900(20) &   1.050(10) & 0.902(15)         \\
 \cline{1-5}
   0.3     & 1.085(10) &    0.879(15)    &     &              \\
%\hline \hline
   0.4  &  1.092(11)   & 0.879(15)    & 1.055(11) & 0.880(12)    \\
\cline{1-5}
  0.5   & 1.060(11)     & 0.867(15) &  &    \\
\cline{1-5}
 0.6 & 1.025 (15) & 0.867(16) &  &   \\
\cline{1-5}
0.7   &   &     &0.943(20) &0.835(16)    \\
\cline{1-5}
0.8   &   &     &0.950(20) &0.890(18)    \\
\hline\hline
\end{tabular}
\caption{Vortex number exponent $\phi_V$ and the potential energy exponent $\phi_U$.}
%\protect\cite {fis88,bra94}.
\label{t2}
\end{table}

%%%%%%%%%%%%%%%%%%%%%%%%%%%%%%%%%%%%%
%We see that the vortex density relaxations
%exponents $\phi_v $ takes values slightly larger than unity.

Now, for a unform distribution of vortices, we can expect the
average separation between the vortices to satisfy (for negligible
$N_V(\infty)$)
\begin{equation} \label{eq:vortex_scaling}
L(t) \sim L_V (t) \sim (N_V (t))^{-1/2} \sim t^{-\phi_V /2}.
\end{equation}
Since we have $L(t) \sim t^{1/z}$, we expect $1/z = \phi_V /2 $. 
The two exponents $1/z $ and $\phi_V /2 $
obtained from simulations satisfy this relation only approximately.
To be more precise, systematically a small discrepancy exists
such that $1/z $ is slightly larger than $\phi_V /2 $.

%%\bigskip
%%   [[[ FIGURE 3 ]]]
%%\bigskip

We find also that the relaxation of the potential energy exhibits a
power law decay $ \Delta U \equiv U(t)-U_{\infty} \sim t^{-\phi_U}$
toward a long-time equilibrium value, with the exponents taking values
approximately in the range $ 0.86 < \phi_U  < 0.90$ as shown in
Fig.~3(b) and Fig.~3(c). The value of this exponent is almost
independent of the energy $E$. For the usual dissipative
coarsening with a non-conservative order parameter in two dimensions,
the excess energy relaxation can be related to the growing length scale
$L(t)$ as
\begin{equation}
\Delta U \sim L^{-2} (t) \log (L(t)/r_{0}),
\end{equation}
where $r_{0}$ denotes the size of a vortex core  \cite{bray-ruten}.
Interestingly, in our model system with Hamiltonian dynamics, we
could not show confidently that the simulation results are consistent
with the above relation  between the energy and the domain growth
relation. Further work is necessary in this regard \cite{hamil_xy_2}. 

Another interesting feature of the coarsening dynamics in
the Hamiltonian {\em XY} model is that the spin autocorrelation exhibits
a peculiar non-power law behavior that can be reasonably well
fitted by 
\begin{equation} \label{eq: Auto_correlation}
A(t) \sim  A_0 \exp [ -b (\ln (t) )^{\gamma} ],
\end{equation}
with $b$ ranging from $0.24$ to $0.4$ and the exponent $\gamma $
from $1.5$ to $1.7$. One typical example is shown in Fig.~4 for the
case of $E=0.4$ with $L=1800$, which shows an excellent fit to the
above functional form with
 $b \simeq 0.242$ and $\gamma \simeq 1.67$. Note that, in the limit of
 $\gamma = 1$, $A(t)$ reduces to a power law of $ A(t) \sim  A_0 t^{-b}$.
The fitted values of $\gamma $ (ranging from $1.5$ to $1.7$) implies   
that $A(t)$ exhibits a considerable deviation from a power-law behavior.
To the best of our knowledge, the Hamiltonian {\em XY} model on a square lattice
is the first system in which the autocorrelation function exhibits
 a functional form with an exponential of a logarithmic power.
%%\bigskip
%%   [[[ FIGURE 4 ]]]
%%\bigskip

Since we have no theories available on the coarsening of the Hamiltonian
{\em XY} model in two dimensions, we may resort to phenomenological
approaches for the explanation of these results on the phase
ordering of the Hamiltonian {\em XY} model, especially the relationship
between the characteristic growth exponents and other exponents obtained from
our simulations.

To begin with, let us note that a strong contrast exists between the
above results of Hamiltonian coarsening and the conventional dissipative
coarsening (via Monte Carlo or Langevin dynamics) in the {\em XY} model.
The first thing to recall is that the growth law of the usual {\em XY} model 
with Monte Carlo dynamics (on nonconserved order parameter) follows 
$L(t) \sim (t/\ln(t))^{1/2}$.
In contrast, here, we observe in the Hamiltonian {\em XY}
model, a growth law with the exponent $1/z$ being greater than the exponent
for the diffusive growth of $1/z=1/2$. Also, the vortex number density
exhibits a relaxation $N_V - N_V(\infty) \sim t^{-\phi_V}$ with
$\phi_V$ greater than $1.0$.

As we mentioned earlier, the system is expected to decrease in its
total potential energy via a corresponding increase of kinetic energy.
This kinetic energy part is thought to consist of spin wave oscillations
(superposed upon vortices or anti-vortices). We expect growth laws of 
length scales to be determined by the effective dynamics of vortices and
anti-vortices and by the statistics of vortex-antivortex annihilation. 

%Since we do not have an available
%theory of full microscopic dynamics of multi-vortex states, we can
%focus on a phenomenological approach on the effective diffusive
%dynamics of a single vortex under an external force (due to another
%vortex, e.g.)

Even though our model system is very simple in its equations, we
find it not very easy to find an effective vortex dynamics for a
given total energy of the system. One way may be to find a Langevin-type
model system, which is expected to be equivalent to our
microcanonical Hamiltonian system in the thermodynamic limit, and
then to investigate the coarsening dynamics of the equivalent model.

Noting that the equation of motion can be derived from the
Hamiltonian
\begin{equation} \label{eq:Hamiltonian2}
H  =  \sum_{<ij>}(1-\cos(\theta_i - \theta_j)) + \sum_i \frac{m_i^2}{2}
\end{equation}
with the Poisson bracket $\{ \theta_i , m_j \} = g$ (where $g=1$ here) and 
considering that the total sum of $m_i$, i.e., $\sum_i m_i$, is conserved,
we can construct a Langevin equation for $\theta$ and $m$ in the following
way: We suppose, to begin with, that the fluctuating spin waves act as
a thermal bath on the vortices, with the temperature determined by the
equivalent equilibrium temperature corresponding to the total energy.
Thus, we may suppose the effective coarsening dynamics of the
original Hamiltonian {\em XY} model system to be equivalent to the
dissipative dynamics of the system under thermal noise associated
with the equivalent temperature:

\begin{eqnarray} \label{eq:mct_xy_eq}
\dot{\theta}_i  & =  &  g\frac{\delta H }{\delta m_i}-\lambda
\frac{\delta H }
 {\delta \theta_i} + \eta_i (t), \\
 \dot{m}_i & = &  -g\frac{\delta H }{\delta \theta_i}+\mu \nabla^2 \frac{\delta H }
 {\delta m_i} + \xi_i (t), 
\end{eqnarray}
where the Gaussian noises $\eta_i$ and $\xi_i$ satisfy
\begin{eqnarray}
\langle \eta_{i}(t) \eta_{j} (t^{\prime}) \rangle & = & 2
\lambda k_B T \delta_{ij} \delta (t-t^{\prime}), \\
\langle \xi_{i}(t) \xi_{j} (t^{\prime}) \rangle & = & 2 \mu k_B T  \
\nabla^2 \delta_{ij} \delta (t-t^{\prime}) 
\end{eqnarray}
at temperature $T$, $k_B $ being the Boltzmann constant. 
Here, $\delta_{ij} $ represents the discrete delta function on
a two-dimensional square lattice, and $\nabla^2$ the corresponding discrete
Lapacian reflecting the conservation of the total {\em Josephson} angular momentum,
$m$. 
In the linear limit, ignoring the vortices, these set of equations 
reduce to the Nelson-Fisher model describing low-temperature spin
dynamics in one and two dimensions \cite{nelson-fisher}.   

It is not a trivial matter to prove an equivalence between the
relaxation of the Hamiltonian {\em XY} model and that of the above 
Langevin equations with reversible mode-coupling. In order to be sure of 
the equivalence, we should assume that the spin dynamics is sufficiently 
chaotic so that the spin-wave fluctuations generate fully Gaussian random
noise on the spin dynamics. Beginning with Eqs.~14 and 15, we may
also obtain an effective dynamics of a single-vortex excitation under external
force.

By assuming a slowly moving vortex, we attempted to apply a
collective variable method \cite{bishop,kamppeter1,kamppeter2} to
Eqs.~14 and 15 in order to obtain an effective equation of
motion for a single vortex. We obtain from this method an
interesting result that, due to the conservation of $m$, the
effective equation for the slow-velocity limit
 of a vortex can be written as
\begin{equation} \label{eq:vortex-dyn}
M \dot{\vec{V}} + \zeta \vec{V} = \vec{F} + \vec{F}_{noise},
\end{equation}
with the effective  mass $M$ of a vortex behaving as $M \sim \ln(L)$, 
 and the noise term $\vec{F}_{noise}$ being an uncorrelated Gaussian
noise. $\zeta$ represents the effective linear friction constant of 
the vortex. $\vec{F}$ is an external force, such as coulomb forces due to
other vortices, etc. An interesting feature is that the dissipation
term vanishes to the lowest order in the velocity of the vortex 
i.e., $\zeta \sim 0 + c|V| + \cdots $.
Including the next order terms is a complicated task that we didn't
actually carry out, but we can attempt a phenomenological argument on
the coarsening by assuming a higher-order dissipation. Here, let us assume
a dissipation term that is second order in the velocity of a vortex,
and let us consider a vortex-antivortex pair with the coulomb force
varying inversely in the mutual distance (in two dimensions). Then,
a simple argument based on a force balance between the coulomb force and 
friction leads to (neglecting the inertial term in the limit of 
slow velocity and at low temperatures)  
\begin{equation} \label{eq:force-balance}
 c V^2 = c({dr \over dt})^2 \sim {1\over r}.   
\end{equation}
In this case, the time scale for annihilation of a vortex-antivortex pair
with size $R$ goes as $\tau (R) \sim R^{3/2} $. This, in turn, implies
that the typical length scale between vortices will grow in time as
$L(t) \sim t^{2/3}$, which is definitely superdiffusive. This value of 
the growth exponent, $2/3$, is a little larger than those exponents obtained
from our simulations. This discrepancy might be attributable to the effect
of vortex inertia, which diverges logarithmically in the vortex size.          
Further work is necessary for a detailed  quantitative understanding 
of the growth laws.  

% conclusion that the growth exponent falls in
%the range $1/2 < 1/z <1 $
%depending on the dominant exponent value of the nonlinear relation
%between dissipation and the vortex velocity. 
%We should assume that the spin dynamics is sufficiently chaotic so that the spin wave
%fluctuations generate fully gaussian random noise on the spin dynamics.

We should note that, in our Hamiltonian dynamics simulations, the initial
states are not completely random states due to the finite energy constraint,
Therefore, the initial states already have some finite correlation length 
scale, i.e., separation $L_c$ between vortices. These length scales are 
of order of $3$ to $5$ lattice constants. Therefore, we expect that  
a finite time scale $\tau_c $ (corresponding to the time scale for a vortex
to travel the distance $L_c$) exists after which a scaling region emerges,
which is well reproduced in our simulations.

\bigskip

\noindent

\medskip

\section{CONCLUSIONS}

We have studied the coarsening dynamics of the Hamiltonian {\em XY}
model on two-dimensional square lattice. An initial state that is
specially tuned to have a given potential energy (otherwise random)
but with zero kinetic energy develops into a late-time coarsening
state, where the potential energy slowly decays as a power law with
a compensating increase in the kinetic energy. We find that the
coarsening dynamics exhibits a characteristic superdiffusive growth,
with the exponents being consistently larger than the value for 
usual diffusive growth of $1/z =1/2$. Relaxation of various quantities cannot be
understood in the usual framework of the dissipative dynamics of the
Ginzburg-Landau Hamiltonian. We believe that these novel features of
coarsening dynamics come from conservation of angular
momentum, which, in turn, causes an effective nonlinear dissipation of
individual vortices.

% the overdamped limit under zero external driving current.  In the case
%f unfrustrated arrays, we find that the friction constant of a
%vortex remains finite in the limit of large extent of the vortex;
%this is argued to lead to the absence of logarithmic factors in
%the length scale and in the vortex number decay, which is in
%reasonable agreement with simulation results. In the case of fully
%frustrated arrays, relaxation and coarsening dynamics are
%characterized by decay of line defects (chirality domain walls) as
%well as of point defects (corner charges). Here strong
%temperature dependence is found in the characteristics of the
%late-time domain growth, which can be explained by activated
%domain wall dynamics across the energy barriers with logarithmic size
%dependence.
%In the early-time regime, on the other handr, there exists a relatively long
%period of transient dynamics which is independent of the temperature;
%this may be attributed to the effects of initial random distributions
%of the strong Lorentz force, making the thermal noise irrelevant.
%It would be of interest to investigate the coarsening dynamics of
%Josephson-junction arrays under external magnetic fields and to compare with
%our simulation results.

\begin{acknowledgments}
%\center{acknowledgments}
This work was supported by Grant R01-2003-000-11595-0 from the Basic 
Research Program of the Korea Science and Engineering Foundation (KOSEF). 
SJL thanks Prof. Jooyoung Lee at the Korea Institute for Advanced Study 
(KIAS), where part of the work was completed, for his hospitality.
\end{acknowledgments}

\newpage

\begin{figure}[fig1]
\includegraphics[width=10.0cm]{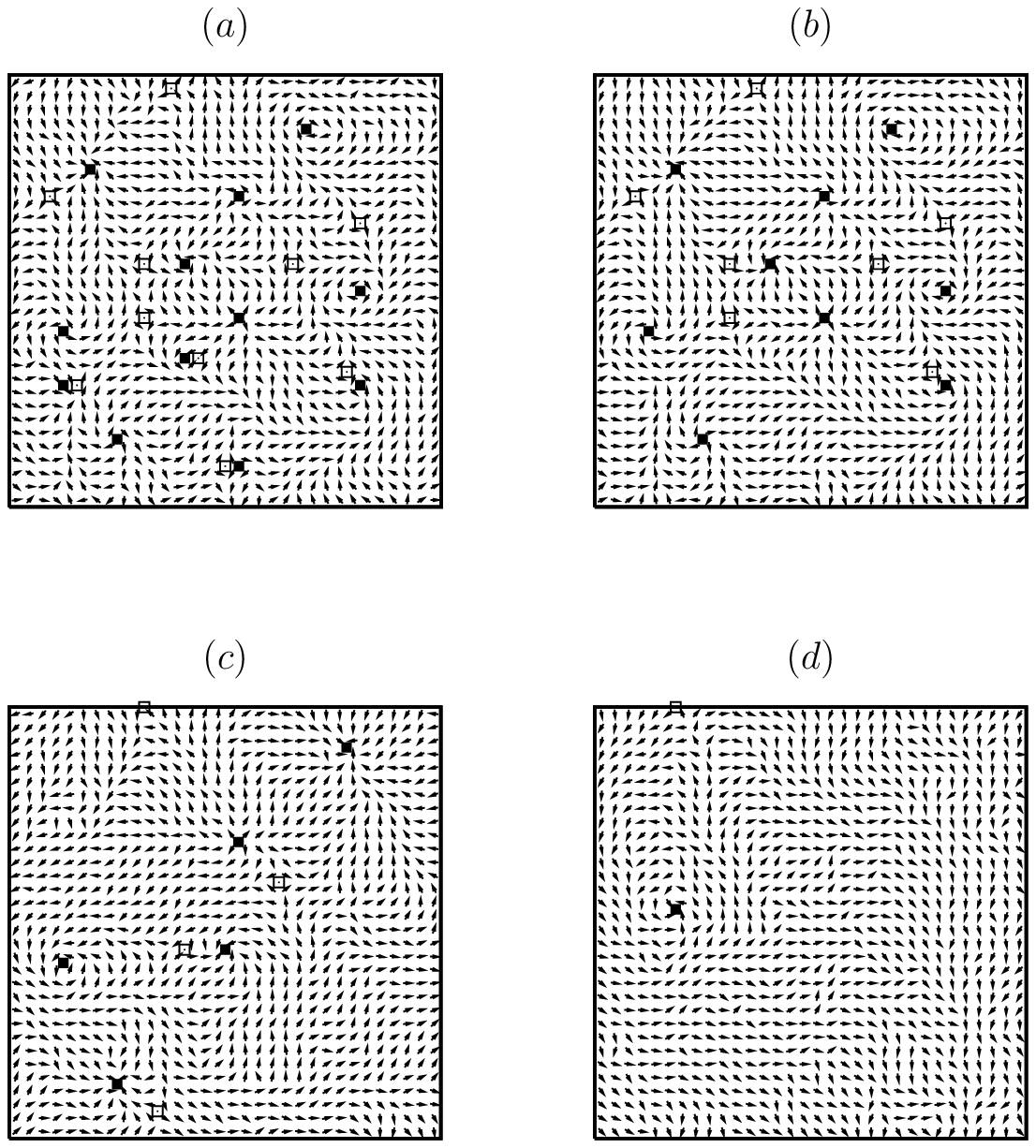}
\caption{
Snapshots of the configuration of the XY spins in a square lattice
of size $L= 32$, $E=0.4$ and time $t$: (a) $t=0$, (b)
$t=0.01 \cdot 2^{7}$, (c) $t=0.01\cdot 2^{10}$, and (d) $t=0.01\cdot 2^{13}$.
Vortices and antivortices are denoted by solid squares and empty
squares, respectively.} \label{fig.1}
\end{figure}

\newpage
\bigskip
\vspace{4.0cm}
\begin{figure}[fig2]
\includegraphics[width=9.0cm]{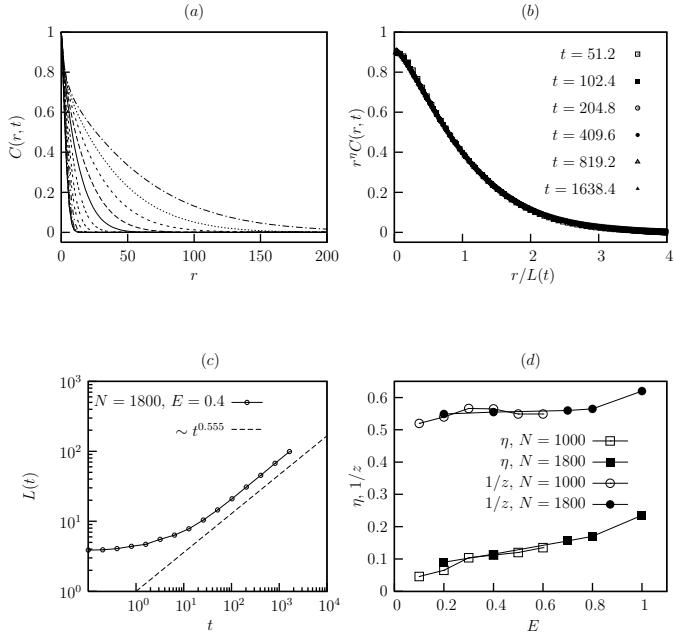}
\caption{
(a) Spatial behavior of the equal time XY spin correlation function
at various time stages in square lattices of size $N = 1800$ at a
per-site energy of $E=0.4$.  (b) Scaling collapse of the data in (a)
with the appropriate scaling length $L(t)$.  (c) Scaling length
versus time at various temperatures $E= 0.4$ and $N = 1800$,
$N=1000$. (d) Growth exponents and $\eta (E)$ versus energy,
manifesting superdiffusive growth. Error bars are of the order of the 
sizes of the symbols or smaller. Lines are only guides to the eyes.
} \label{fig.2}
\end{figure}

\newpage
\bigskip
\vspace{4.0cm}

\begin{figure}[fig3]
\includegraphics[width=9.0cm]{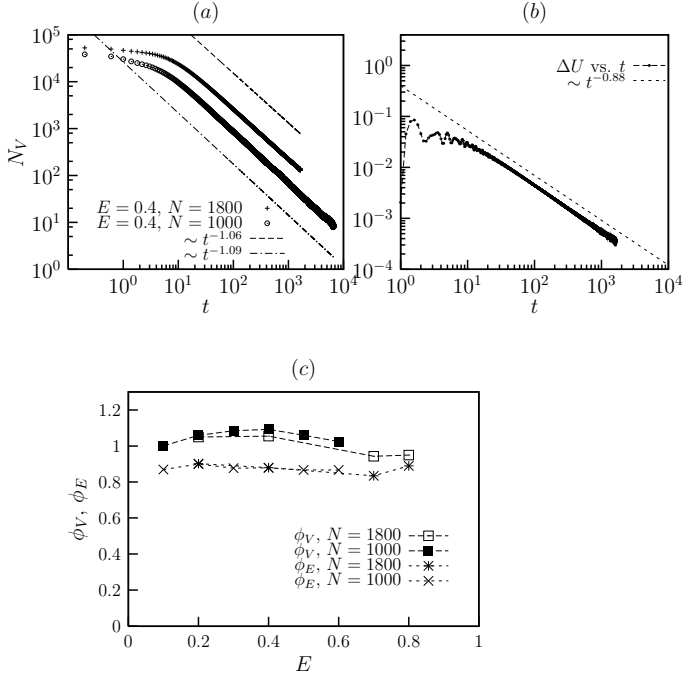}
\caption{
Relaxation of (a) the vortex number $N_v$ at energy $E=0.40$ and
system size $N=1800$ and $1000$ and of (b) the excess potential energy
$\Delta U$ at $E=0.4$ in the case of size $L=1800$. In (a), the
power-law lines  $t^{-1.06}$ and $t^{-1.09}$  are also plotted
whereas in (b) a power law line with $\sim t^{-0.88}$ is shown. (c)
shows the exponents $\phi_V $ and $\phi_U$ versus energy.
Error bars are of the order of the sizes of the symbols. Lines are only 
guides to the eyes.} \label{fig.3}
\end{figure}

\newpage
\bigskip
\vspace{4.0cm}

\begin{figure}[fig4]
\includegraphics[width=9.5cm]{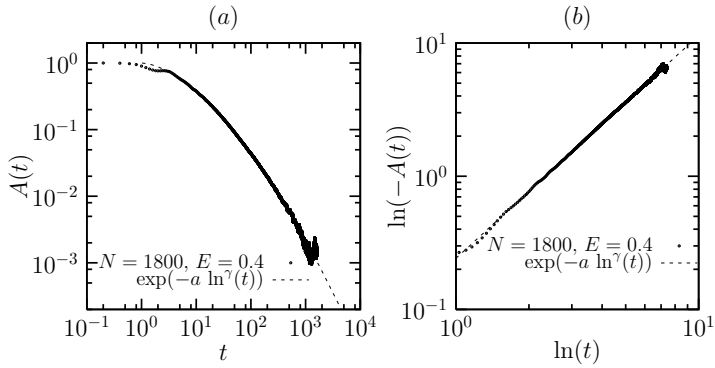}
\caption{
Spin autocorrelation function at energy $E = 0.40$ and (b) the same
in a double logarithmic plot. The fitted values are  
$b\simeq 0.242$ and $\gamma \simeq 1.67$.} \label{fig.4}
\end{figure}

\end{document}